\documentclass[final]{IEEEtran}

\usepackage{setspace} 
\usepackage{multicol}
\usepackage[final]{graphicx}
\usepackage{wrapfig}
\usepackage{bm}
\usepackage{amsmath,amssymb,amsthm}

\usepackage{algorithm}
\usepackage{algorithmic}
\usepackage{bm,amsmath,amssymb,amsthm,graphicx,graphics,subfigure,epsfig,enumerate}
\usepackage{bbm}
\usepackage{amsfonts}

\graphicspath{{graphs/}}
\newcommand{\balpha}{{\bm{\alpha}}}
\newcommand{\bT}{{\mathbf T}}
\newcommand{\bC}{{\mathbf C}}
\newcommand{\bD}{{\mathbf D}}
\newcommand{\bd}{{\mathbf d}}
\newcommand{\bv}{{\mathbf v}}

\newcommand{\bG}{{\mathbf G}}
\newcommand{\bn}{{\mathbf n}}
\newcommand{\ba}{{\mathbf a}}

\newcommand{\bx}{{\mathbf x}}
\newcommand{\bU}{{\mathbf U}}
\newcommand{\bQ}{{\mathbf Q}}
\newcommand{\by}{{\mathbf y}}
\newcommand{\br}{{\mathbf r}}

\newcommand{\bR}{{\mathbf R}}
\newcommand{\bX}{{\mathbf X}}
\newcommand{\bH}{{\mathbf H}}
\newcommand{\bV}{{\mathbf V}}
\newcommand{\bW}{{\mathbf W}}

\newcommand{\bs}{{\mathbf s}}
\newcommand{\bY}{{\mathbf Y}}
\newcommand{\bP}{{\mathbf P}}

\newcommand{\bp}{{\mathbf p}}
\newcommand{\bI}{{\mathbf I}}
\newcommand{\bZ}{{\mathbf Z}}

\newcommand{\cP}{{\mathcal P}}

\newcommand{\cN}{{\mathcal N}}

\newcommand{\argmin}{\mathop{\rm argmin}}

\title{PETRELS: Parallel Subspace Estimation and Tracking by Recursive Least Squares from Partial Observations}
\author{Yuejie Chi$^*$, Yonina C. Eldar and Robert Calderbank \\
\thanks{Y. Chi is with the Department of Electrical and Computer Engineering, The Ohio State University,
Columbus, OH 43210, USA (email: chi@ece.osu.edu).}
\thanks{Y. C. Eldar is with the Department of Electrical Engineering, Technion, Israel Institute of Technology, Haifa 32000, Israel (email: yonina@ee.technion.ac.il).}
\thanks{R. Calderbank is with the Department of Computer Science, Duke University, Durham, NC 27708, USA (email: robert.calderbank@duke.edu). }
\thanks{The work of Y. Chi and R. Calderbank was supported by ONR under Grant N00014-08-1-1110, by AFOSR under Grant FA 9550-09-1-0643, and by NSF under Grants NSF CCF -0915299 and NSF CCF-1017431. The work of Y. Eldar was supported in part by the Ollendorf foundation, and by the Israel Science Foundation under Grant 170/10.}
\thanks{A preliminary version of this work was presented at the 2012 International Conference on Acoustics, Speech, and Signal Processing (ICASSP).}
}
\date{\today}
\usepackage{times} % alt: cmbright

\begin{document}

\maketitle

\begin{abstract}
Many real world datasets exhibit an embedding of low-dimensional structures in a high-dimensional manifold. Examples include images, videos and internet traffic data. It is of great significance to reduce the storage requirements and computational complexity when the data dimension is high. Therefore we consider the problem of reconstructing a data stream from a small subset of its entries, where the data is assumed to lie in a low-dimensional linear subspace, possibly corrupted by noise. We further consider tracking the change of the underlying subspace, which can be applied to applications such as video denoising, network monitoring and anomaly detection. Our setting can be viewed as a sequential low-rank matrix completion problem in which the subspace is learned in an online fashion. The proposed algorithm, dubbed Parallel Estimation and Tracking by REcursive Least Squares (PETRELS), first identifies the underlying low-dimensional subspace, and then reconstructs the missing entries via least-squares estimation if required. Subspace identification is perfermed via a recursive procedure for each row of the subspace matrix in parallel with discounting for previous observations. Numerical examples are provided for direction-of-arrival estimation and matrix completion, comparing PETRELS with state of the art batch algorithms.
\end{abstract}

\section{Introduction}
Many real world datasets exhibit an embedding of low-dimensional structures in a high-dimensional manifold. When the embedding is assumed linear, the underlying low-dimensional structure becomes a linear subspace. Subspace Identification and Tracking (SIT) plays an important role in various signal processing tasks such as online identification of network anomalies \cite{Ahmed_Coates_Lakhina_2007}, moving target localization \cite{Shahbazpanahi_2001}, beamforming \cite{Kumaresan_Tufts_1983}, and denoising \cite{Sayed_2003}. Conventional SIT algorithms collect full measurements of the data stream at each time, and subsequently update the subspace estimate by utilizing the track record of the stream history in different ways \cite{Yang-1995, Crammer_2006}. 

Recent advances in Compressive Sensing (CS) \cite{CT2005, Donoho2006} and Matrix Completion (MC) \cite{Candes-MC,Candes_Recht_2008} have made it possible to infer data structure from highly incomplete observations. Compared with CS, which allows reconstruction of a single vector from only a few attributes by assuming it is sparse in a pre-determined basis or dictionary, MC allows reconstruction of a matrix from a few entries by assuming it is low rank. A popular method to perform MC is to minimize the nuclear norm of the corresponding matrix \cite{Candes-MC,Candes_Recht_2008} that the observed entries are satisfied. This method requires no prior knowledge of rank, in a similar spirit with $\ell_1$ minimization \cite{chen1998atomic} for sparse recovery in CS. Other approaches including greedy algorithms such as OptSpace \cite{OptSpace} and ADMiRA \cite{lee2010admira} require an estimate of the matrix rank for initialization. Identifying the underlying low-rank structure in MC is equivalent to subspace identification in a batch setting. When the number of observed entries is slightly larger than the subspace rank, it has been shown that with high probability, it is possible to test whether a highly incomplete vector of interest lies in a known subspace \cite{Balzano-subspace}. Recent works on covariance matrix and principal components analysis of a dataset with missing entries also validate that it is possible to infer the principal components with high probability \cite{lounici2012high,chi2012rnsc}. 

In high-dimensional problems, it might be expensive and even impossible to collect data from all dimensions. For example in wireless sensor networks, collecting data from all sensors continuously will quickly drain the battery power. Ideally, we would prefer to obtain data from a fixed budget of sensors of each time to increase the overall battery life, and still be able to identify the underlying structure. Another example is in online recommendation systems, where it is impossible to expect rating feedbacks from all users on every product are available. Therefore it is of growing interest to identify and track a low-dimensional subspace from highly incomplete information of a data stream in an online fashion. In this setting, the estimate of the subspace is updated and tracked across time when new observations become available with low computational cost. The GROUSE algorithm \cite{Balzano-2010} has been recently proposed for SIT from online partial observations using rank-one updates of the estimated subspace on the Grassmannian manifold. However, performance is limited by the existence of ``barriers'' in the search path \cite{SET} which result in GROUSE being trapped at a local minima. We demonstrate this behavior through numerical examples in Section~\ref{simulation} in the context of direction-of-arrival estimation.

In this paper we further study the problem of SIT given partial observations from a data stream as in GROUSE. Our proposed algorithm is dubbed Parallel Estimation and Tracking by REcursive Least Squares (PETRELS). The underlying low-dimensional subspace is identified by minimizing the geometrically discounted sum of projection residuals on the observed entries per time index, via a recursive procedure with discounting for each row of the subspace matrix in parallel. The missing entries are then reconstructed via least-squares estimation if required. The discount factor balances the algorithm's ability to capture long term behavior and changes to that behavior to improve adaptivity. We also benefit from the fact that our optimization of the estimated subspace is on all the possible low-rank subspaces, not restricted to the Grassmannian manifold. In the partial observation scenario, PETRELS always converges locally to a stationary point since  it it a second-order stochastic gradient descent algorithm. In the full observation scenario, we prove that PETRELS actually converges to the global optimum by revealing its connection with the well-known Projection Approximation Subspace Tracking (PAST) algorithm \cite{Yang-1995}. Finally, we provide numerical examples to measure the impact of the discount factor, estimated rank and number of observed entries. In the context of direction-of-arrival estimation we demonstrate superior performance of PETRELS over GROUSE in terms of separating close-located modes and tracking changes in the scene. We also compare PETRELS with state of the art batch MC algorithms, showing it as a competitive alternative when the subspace is fixed.
%We further provide extension of the algorithm in terms of further reducing the storage and computational requirements and improving robustness. 

The rest of the paper is organized as follows. Section~\ref{background} states the problem and provides background in the context of matrix completion and conventional subspace tracking. Section~\ref{algorithm} describes the algorithm in details.  Two extensions of PETRELS to improve robustness and reduce complexity are presented in Section~\ref{discussions}. We discuss convergence issues of PETRELS in the full observation scenario in Section~\ref{convergence}. Section~\ref{simulation} shows the numerical results and we conclude the paper in Section~\ref{conclusion}.

\section{Problem Statement and Related Work} \label{background}

\subsection{Problem Statement}
We consider the following problem. At each time $t$, a vector $\bx_t \in\mathbb{R}^M$ is generated as:
\begin{equation} \label{model_data}
\bx_t=\bU_t\ba_t+\bn_t \in\mathbb{R}^M,
\end{equation}
where the columns of $\bU_t\in\mathbb{R}^{M\times r_{t}}$ span a low-dimensional subspace, the vector $\ba_t\in\mathbb{R}^{r_t}$ specifies the linear combination of columns and is Gaussian distributed as $\ba_t\sim\mathcal{N}(\mathbf{0}, \bI_{r_t})$, and $\bn_t$ is an additive white Gaussian noise distributed as $\bn_t\sim\mathcal{N}(0, \sigma^2\bI_M)$. The rank of the underlying subspace $r_t$ is not assumed known exactly and can be slowly changing over time. The entries in the vectors $\bx_t$ can be considered as measurements from different sensors in a sensor network, values of different pixels from a video frame, or movie ratings from each user. 

We assume only partial entries of the full vector $\bx_t$ are observed, given by 
\begin{equation} 
\by_t=\bp_t\odot \bx_t=\bP_t\bx_t \in\mathbb{R}^{M},
\end{equation}
where $\odot$ denotes point-wise multiplication, $\bP_t=\mbox{diag}\{\bp_t\}$, $\bp_t=[p_{1t}, p_{2t}, \cdots, p_{Mt}]^T\in\{0,1\}^{M}$ with $p_{mt}=1$ if the $m$th entry is observed at time $t$. We denote $\Omega_t=\{m: p_{mt}=1\}$ as the set of observed entries at time $t$. In a random observation model, we assume the measurements are taken uniformly at random. 
%each $p_{mt}$ is a  Bernoulli random variable with $\Pr(p_{mt}=1)=p$ and $\Pr(p_{mt}=0)=1-p$.

We are interested in an online estimate of a low-rank subspace $\bD_n\in\mathbb{R}^{M\times r}$ at each time index $n$, which identifies and tracks the changes in the underlying subspace, from streaming partial observations $(\by_t,\bp_t)_{t=1}^n$. The rank of the estimated subspace $\bD_n$ is assumed known and fixed throughout the algorithm as $r$. In practice, we assume the upper bound of the rank of the underlying subspace $\bU_t$ is known, and $\sup_t r_t \leq r$. The desired properties for the algorithm include:
\begin{itemize}
\item \textit{Low complexity:} each step of the online algorithm at time index $n$ should be adaptive with small complexity compared to running a batch algorithm using history data;
\item \textit{Small storage:} The online algorithm should require a storage size that does not grow with the data size;
\item \textit{Convergence:} The subspace sequence generated by the online algorithm should converge to the true subspace $\bU_t=\bU$ if it is constant.
\item \textit{Adaptivity:} The online algorithm should be able to track the changes of the underlying subspace in a timely fashion.
\end{itemize}

\subsection{Conventional Subspace Identification and Tracking}
When $\bx_t$'s are fully observed, our problem is equivalent to the classical SIT problem, which is widely studied and has a rich literature in the signal processing community. Here we describe the Projection Approximation Subspace Tracking (PAST) algorithm in details which is the closest to our proposed algorithm in the conventional scenario. 

First, consider optimizing the scalar function with respect to a subspace $\bW\in\mathbb{R}^{M\times r}$, given by
\begin{equation}\label{past-opt}
J(\bW) = \mathbb{E}\|\bx_t-\bW\bW^T\bx_t\|_2^2.
\end{equation}
When $\bU_t=\bU$ is fixed over time, let  $\bC_{\bx}=\mathbb{E}[\bx_t\bx_t^T]=\bU\bU^T+\sigma^2 \bI_M$ be the data covariance matrix. It is shown in \cite{Yang-1995} that the global minima of \eqref{past-opt} is the only stable stationary point, and is given by $\bW=\bU_r\bQ$, where $\bU_r$ is composed of the $r$ dominant eigenvectors of $\bC_{\bx}$, and $\bQ\in\mathbb{C}^{r\times r}$ is a unitary matrix. Without loss of generality, we can choose $\bU_r=\bU$. This motivates PAST to optimize the following function at time $n$ without constraining $\bW$ to have orthogonal columns:
\begin{align}
\bW_n &= \argmin_{\bW\in\mathbb{R}^{M\times r}} \sum_{t=1}^n \alpha^{n-t} \|\bx_t-\bW\bW^T\bx_t\|_2^2,\label{exact-past} \\
&\approx \argmin_{\bW\in\mathbb{R}^{M\times r}} \sum_{t=1}^n \alpha^{n-t} \|\bx_t-\bW\bW_{n-1}^T\bx_t\|_2^2, \label{approx-past}
\end{align}
where the expectation in \eqref{past-opt} is replaced by geometrically reweighting the previous observations by $\alpha$ in \eqref{exact-past}, and is further approximated by replacing the second $\bW$ by its previous estimate in \eqref{approx-past}. Based on \eqref{approx-past}, the subspace $\bW_n$ can be found by first estimating the coefficient vector $\ba_n$ using the previous subspace estimate as $\ba_n=\bW_{n-1}^T\bx_n$, and updating the subspace as
\begin{align}
\bW_n &= \argmin_{\bW\in\mathbb{R}^{M\times r}} \sum_{t=1}^n \alpha^{n-t} \|\bx_t-\bW\ba_t\|_2^2. 
\end{align}

Suppose that $\alpha=1$ and denote $\bR_n=\sum_{i=1}^n \ba_n\ba_n^T$. In \cite{Yang-1996}, the asymptotic dynamics of the PAST algorithm is described by its equilibrium as time goes to infinity using the Ordinary Differential Equation (ODE) below:
\begin{align*}
\dot{\bR} & = \mathbb{E}[\tilde{\ba}_n\tilde{\ba}_n^T] -\bR = \bW^T\bC_{\bx}\bW - \bR, \\
\dot{\bW} & = \mathbb{E}[\bx_n(\bx_n-\bW\tilde{\ba}_n)^T] \bR^{\dag} = (\bI - \bW\bW^T)\bC_{\bx}\bW\bR^{\dag}, 
\end{align*}
where $\tilde{\ba}_n = \bW^T \bx_n$, $\bR=\bR(t)$ and $\bW=\bW(t)$ are continuous time versions of $\bR_n$ and $\bW_n$, and $\dag$ denotes the pseudo-inverse. It is proved in \cite{Yang-1996} that as $t$ increases, $\bW(t)$ converges to the global optima, i.e. to a matrix which spans the eigenvectors of $\bC_{\bx}$ corresponding to the $r$ largest eigenvalues. In Section \ref{convergence} we show that our proposed PETRELS algorithm becomes essentially equivalent to PAST when all entries of the data stream are observed, and can be shown to converge globally.

The PAST algorithm belongs to the class of power-based techniques, which include the Oja's method \cite{Oja_1982}, the Novel Information Criterion (NIC) method \cite{Miao_1998} and etc: These algorithms are treated under a unified framework in \cite{Hua_1999} with slight variations for each algorithm. The readers are referred to \cite{Hua_1999} for details. In general, the estimate of the low-rank subspace $\bW_n\in\mathbb{R}^{M\times r}$ is updated at time $n$ as
\begin{equation} \label{PowerM}
\bW_{n} = \bC_{n} \bW_{n-1}(\bW_{n-1}^T\bC^2_{n}\bW_{n-1})^{-1/2},
\end{equation}
where $\bC_{n}$ is the sample data covariance matrix updated from
\begin{equation} \label{update_covariance}
\bC_n = \alpha_n \bC_{n-1} +\bx_n\bx_n^T,
\end{equation}
and $\alpha_n$ is a parameter between $0$ and $1$. The normalization in \eqref{PowerM} assures that the updated subspace $\bW_n$ is orthogonal but this normalization is not performed strictly in different algorithms. 

It is shown in \cite{Hua_1999} that these power-based methods guarantee global convergence to the principal subspace spanned by eigenvectors corresponding to the $r$ largest eigenvalues of $\bC_{\bx}$. If the entries of the data vector $\bx_t$'s are fully observed, then $\bC_n$ converges to $\bC_{\bx}$ very fast, and this is exactly why the power-based methods perform very well in practice. When the data is highly incomplete, the convergence of \eqref{update_covariance} is very slow since only a small fraction $|\Omega_n|^2/n^2$ of entries in $\bC_{n-1}$ are updated, where $|\Omega_n|$ is the number of observed entries at time $n$, making direct adoption of the above method unrealistic in the partial observation scenario.

\subsection{Matrix Completion}
When only partial observations are available and $\bU_t=\bU$ are fixed, our problem is closely related to the Matrix Completion (MC) problem, which has been extensively studied recently. Assume $\bX\in\mathbb{R}^{M\times n}$ is a low-rank matrix, $\bP$ is a binary $M\times n$ mask matrix with $0$ at missing entries and $1$ at observed entries. Let $\bY=\bP\odot \bX=[\by_1,\ldots, \by_n]$ be the observed partial matrix where the missing entries are filled in as zero, and $\odot$ denotes point-wise multiplication. MC aims to solve the following problem:
\begin{equation} \label{rankmin}
\min_{\bZ}\;\;\mbox{rank}(\bZ)~\quad \mbox{s.t.}\; \; \bY-\bP\odot \bZ=\mathbf{0},
\end{equation}
i.e. to find a matrix with the minimal rank such that the observed entries are satisfied. This problem is combinatorially intractable due to the rank constraint. 

It has been shown in \cite{Candes-MC} that by replacing the rank constraint with nuclear norm minimization, \eqref{rankmin} can be solved by a convex optimization problem, resulting in the following spectral-regularized MC problem:
\begin{equation} \label{MC}
\min_{\bZ} \;\; \frac{1}{2}\| \bY-\bP\odot \bZ \|_F^2 +\mu\|\bZ\|_{*},
\end{equation}
where $\|\bZ\|_{*}$ is the nuclear norm of $\bZ$, i.e. the sum of singular values of $\bZ$, and $\mu>0$ is a regularization parameter. Under mild conditions, the solution of \eqref{MC} is the same as that of \eqref{rankmin} \cite{Candes-MC}. The nuclear norm \cite{Tibshirani_2009} of $\bZ$ is given by
\begin{equation} \label{def_nn}
\|\bZ\|_{*} = \min_{\bU,\bV: \bZ=\bU\bV^T} \frac{1}{2} \left( \|\bU\|_F^2+\|\bV\|_F^2 \right)
\end{equation}
where $\bU\in\mathbb{C}^{M\times r}$ and $\bV\in\mathbb{C}^{n\times r}$. Substituting \eqref{def_nn} in \eqref{MC} we can rewrite the MC problem as 
\begin{equation} \label{MF}
\min_{\bU,\bV} \|\bP\odot (\bX-\bU\bV)\|_F^2 +\mu\left( \|\bU\|_F^2+\|\bV\|_F^2 \right).
\end{equation}
%A regularized version of PETRELS resembles \eqref{MF} as shown in Section~\ref{mcsimilar}.

Our problem formulation can be viewed as an online way of solving the above batch-setting MC problem. Consider a random process $\{n_t\}$ where each $n_t$ is drawn uniformly from $\{1,\ldots, n\}$, and a data stream is constructed where the data at each time is given as $\bx_{n_t}$, i.e. the $n_t$th column of $\bX$. Compared with \eqref{model_data}, the subspace is fixed as $\bU_t=\bU$ since we draw columns from a fixed low-rank matrix. Each time we only observe partial entries of $\bx_{n_t}$, given as $\by_{n_t}=\bp_{n_t}\odot \bx_{n_t}$, where $\bP_{n_t}$ is the $n_t$th column of $\bP$. The problem of MC becomes equivalent to retrieving the underlying subspace $\bU$ from the data stream $(\by_{n_t},\bp_{n_t})_{t=1}^{\infty}$. After estimating $\bU$, the low-rank matrix $\bX$ can be recovered via least-squares estimation. The online treatment of the batch MC problem has potential advantages for avoiding large matrix manipulations. We will compare the PETRELS algorithm against some of the popular MC methods in Section~\ref{simulation}.

\section{The PETRELS Algorithm} \label{algorithm}
We now describe our proposed Parallel Estimation and Tracking by REcursive Least Squares (PETRELS) algorithm.
\subsection{Objective Function}
We first define the function $f_t(\bD)$ at each time $t=1,\cdots, n$ for a fixed subspace $\bD\in\mathbb{R}^{M\times r}$, which is the total projection residual on the observed entries, 
\begin{align} \label{ftD}
f_t(\bD) &= \min_{\ba_t} \|\bP_t(\bx_t-\bD\ba_t)\|_2^2,\;t=1,\cdots, n.
\end{align}
Here $r$ is the rank of the estimated subspace, which is assumed known and fixed throughout the algorithm\footnote{The rank may not equal the true subspace dimension.}. We aim to minimize the following loss function at each time $n$ with respect to the underlying subspace:
\begin{equation}\label{optimization}
\bD_n =  \argmin_{\bD\in\mathbb{R}^{M\times r}} F_n(\bD)= \argmin_{\bD\in\mathbb{R}^{M\times r}} \sum_{t=1}^n \lambda^{n-t}f_t(\bD),
\end{equation}
where $\bD_n$ is the estimated subspace of rank $r$ at time $n$, and the parameter $0\ll \lambda\leq 1$ discounts past observations.

Before developing PETRELS we note that if there are further constraints on the coefficients $\ba_t$'s, a regularization term can be incorporated as:
\begin{equation}\label{regsample}
f_t(\bD) = \min_{\ba_t\in\mathbb{R}^r} \|\bP_t(\bD\ba_t-\bx_t)\|_2^2+\beta \|\ba_t\|_p,
\end{equation}
where $p\geq 0$. For example, $p=1$ enforces a sparse constraint on $\ba_t$, and $p=2$ enforces a norm constraint on $\ba_t$. 

In \eqref{optimization} the discount factor $\lambda$ is fixed, and the influence of past estimates decreases geometrically; a more general online objective function can be given as
\begin{equation}\label{objective}
 F_n(\bD)=\lambda_n F_{n-1}(\bD)+  f_n(\bD),
\end{equation}
where the sequence $\{\lambda_n\}$ is used to control the memory and adaptivity of the system in a more flexible way.

To motivate the loss function in \eqref{optimization} we note that if $\bU_t=\bU$ is not changing over time, then the RHS of \eqref{optimization} is minimized to zero when $\bD_n$ spans the subspace defined by $\bU$. If $\bU_t$ is slowly changing, then $\lambda$ is used to control the memory of the system and maintain tracking ability at time $n$. For example, by using $\lambda\to 1$ the algorithm gradually loses its ability to forget the past. 

Fixing $\bD$, $f_t(\bD)$ can be written as
\begin{align}
f_t(\bD)& =  \bx_t^T\left( \bP_t -\bP_t\bD(\bD^T\bP_t\bD)^{\dag} \bD^T\bP_t \right)\bx_t.
\end{align}
Plugging this back to \eqref{optimization} the exact optimization problem becomes:
\begin{equation*}
\bD_n = \argmin_{\bD\in\mathbb{R}^{M\times r}} \sum_{t=1}^n \lambda^{n-t} \bx_t^T\left[ \bP_t -\bP_t\bD(\bD^T\bP_t\bD)^{\dag} \bD^T\bP_t \right]\bx_t.
\end{equation*}
This problem is difficult to solve over $\bD$ and requires storing all previous observations. Instead, we propose PETRELS to approximately solve this optimization problem.
%Assume that there exists $\bD^*$ such that all $F_\infty(\bD)$ is minimized, the goal for the online algorithm is not to find exact solution for each optimization problem at time $t$, but to find a sequence of $\bD(t)$ that converges to $\bD^*$. %\cite{Letton}

\begin{algorithm}
\caption{PETRELS for SIT from Partial Observations}
\textbf{Input:}  a stream of vectors $\by_t$ and observed pattern $\bP_t$. \\
\textbf{Initialization:}  an $M\times r$ random matrix $\bD_0$, and $(\bR_m^0)^{\dag}=\delta\bI_r$, $\delta>0$ for all $m=1, \cdots, M$.

\begin{algorithmic}[1]
\FOR{$n=1, 2, \cdots$}
\STATE $\ba_n =(\bD_{n-1}^T\bP_n\bD_{n-1})^{\dag}\bD_{n-1}^T\by_n. $
\STATE $\hat{\bx}_n =\bD_{n-1}\ba_n$.
\FOR{$m=1, \cdots, M$}
\STATE $\beta_m^n =1+\lambda^{-1}\ba_n^T(\bR_m^{n-1})^{\dag}\ba_n, $
\STATE $\bv_m^n =\lambda^{-1}(\bR_m^{n-1})^{\dag}\ba_n,$
\STATE  $(\bR_m^n)^{\dag} =\lambda^{-1}(\bR_m^{n-1})^{\dag} - p_{mt} (\beta_m^n)^{-1} \bv_m^n(\bv_m^n)^T,$ 
\STATE $\bd_m^n =\bd_m^{n-1}+p_{mn}(x_{mn}-\ba_n^T\bd_m^{n-1})(\bR_m^n)^{\dag}\ba_n. $
\ENDFOR
\ENDFOR
\end{algorithmic}
\label{PETRELS}
\end{algorithm}

\subsection{PETRELS}
The proposed PETRELS algorithm, as summarized by Algorithm~\ref{PETRELS}, alternates between coefficient estimation and subspace update at each time $n$. In particular, the coefficient vector $\ba_n$ is estimated by minimizing the projection residual on the previous subspace estimate $\bD_{n-1}$:
\begin{align} \label{coef}
\ba_n & =\argmin_{\ba\in\mathbb{R}^r} \|\bP_n(\bx_n-\bD_{n-1}\ba)\|_2^2 \nonumber \\
&= (\bD_{n-1}^T\bP_n\bD_{n-1})^{\dag}\bD_{n-1}^T\by_n,
\end{align}
where $\bD_0$ is a random subspace initialization. The full vector $\bx_n$ is then estimated as:
\begin{align} \label{data_recon}
\hat{\bx}_n &=\bD_{n-1}\ba_n.
\end{align}
The subspace $\bD_n$ is then updated by minimizing
\begin{equation} \label{parD}
\bD_n =\argmin_{\bD}\sum_{t=1}^n\lambda^{n-t}\|\bP_t (\bx_t-\bD\ba_t)\|_2^2,
\end{equation}
where $\ba_t$, $t=1, \cdots, n$ are estimates from \eqref{coef}. Comparing \eqref{parD} with \eqref{optimization}, the optimal coefficients are substituted for the previous estimated coefficients. This results in a simpler problem for finding $\bD_n$. The discount factor mitigates the error propagation and compensates for the fact that we used the previous coefficients updated rather than solving \eqref{optimization} directly, therefore improving the performance of the algorithm. 

The objective function in \eqref{parD} can be equivalently decomposed into a set of smaller problems for each row of $\bD_n=[\bd_1^n,\bd_2^n,\cdots,\bd_M^n]^T$ as
\begin{align} \label{parallel_dm}
\bd_m^n &=\argmin_{\bd_m}\sum_{t=1}^n\lambda^{n-t}p_{mt} (x_{mt}-\ba_t^T\bd_m)^2, 
\end{align} 
for $m=1,\cdots, M$. To find the optimal $\bd_m^n$, we equate the derivative of \eqref{parallel_dm} to zero, resulting in
\begin{align}
\left( \sum_{t=1}^n \lambda^{n-t}p_{mt}\ba_t\ba_t^T\right)\bd_m^n &-\sum_{t=1}^n \lambda^{n-t}p_{mt}x_{mt}\ba_t = \mathbf{0}. \nonumber
\end{align}
This equation can be rewritten as 
\begin{equation}\label{rec-dn}
\bR_m^n \bd_m^n = \bs_m^n, 
\end{equation}
where $\bR_m^n=\sum_{t=1}^n \lambda^{n-t}p_{mt}\ba_t\ba_t^T$ and $\bs_m^n=\sum_{t=1}^n \lambda^{n-t}p_{mt}x_{mt}\ba_t$. Therefore, $\bd_m^n$ can be found as
\begin{align} \label{eq23}
\bd_m^n & = ( \bR_m^n)^{\dag} \bs_m^n.
\end{align}
When $\bR_m^n$ is not invertible, \eqref{eq23} is the least-norm solution to $\bd_m^n$. 

We now show how \eqref{rec-dn} can be updated recursively. First we rewrite
 \begin{align}
 \bR_m^n &=\lambda\bR_m^{n-1}+p_{mn}\ba_n\ba_n^T, \label{rec-R} \\
\quad \bs_m^n &=\lambda\bs_m^{n-1}+p_{mn}x_{mn}\ba_n, \label{rec-s}
 \end{align}
for all $m=1,\cdots, M$. Then we plug \eqref{rec-R} and \eqref{rec-s} into \eqref{rec-dn}, and get 
\begin{align}
\bR_m^n \bd_m^n & = \lambda\bs_m^{n-1} + p_{mn}x_{mn}\ba_n \nonumber \\
& = \lambda\bR_m^{n-1}\bd_m^{n-1} + p_{mn}x_{mn}\ba_n \nonumber \\
& = \bR_m^n\bd_m^{n-1} -p_{mn}\ba_n\ba_n^T\bd_m^{n-1}+ p_{mn}x_{mn}\ba_n  \nonumber \\
&= \bR_m^n \bd_m^{n-1}+p_{mn}(x_{mn}-\ba_n^T\bd_m^{n-1})\ba_n, \label{row-update} 
\end{align} 
where $\bd_m^{n-1}$ is the row estimate in the previous time $n-1$. This results in a parallel procedure to update all rows of the subspace matrix $\bD_n$, give as
 \begin{align}
\bd_{m}^n&= \bd_m^{n-1}+p_{mn}(x_{mn}-\ba_n^T\bd_m^{n-1})(\bR_m^n)^{\dag}\ba_n. \label{row-update} 
\end{align} 
Finally, by the Recursive Least-Squares (RLS) updating formula for the general pseudo-inverse matrix \cite{Golub_Van_Loan_1996, general_PI}, $(\bR_m^n)^\dag$ can be easily updated without matrix inversion using
\begin{align}
(\bR_m^n)^{\dag} & =(\lambda\bR_m^{n-1}+p_{mn}\ba_n\ba_n^T)^{\dag} \nonumber \\
&=\lambda^{-1}(\bR_m^{n-1})^{\dag} - p_{mt}\bG_m^n.
\end{align}
Here $\bG_m^n=  (\beta_m^n)^{-1} \bv_m^n(\bv_m^n)^T$, with $\beta_m^n$ and $\bv_m^n$ given as
\begin{align*} 
\beta_m^n &=1+\lambda^{-1}\ba_n^T(\bR_m^{n-1})^{\dag}\ba_n, \\
\bv_m^n &=\lambda^{-1}(\bR_m^{n-1})^{\dag}\ba_n.
\end{align*}

To enable the RLS procedure, the matrix $(\bR_m^0)^{\dag}$ is initialized as a matrix with large entries on the diagonal, which we choose arbitrarily as  the identity matrix $(\bR_m^0)^{\dag}=\delta\bI_r$, $\delta>0$ for all $m=1, \cdots, M$. It is worth noting that implementation of the fast RLS update rules is in general very efficient. However, caution needs to be taken since direct application of fast RLS algorithms suffer from numerical instability of finite-precision operations when running for a long time \cite{Cioffi_1987}.

\subsection{Second-Order Stochastic Gradient Descent} \label{second-order}
The PETRELS algorithm can be regarded as a second-order stochastic gradient descent method to solve \eqref{optimization} by using $\bd_m^{n-1}$, $m=1,\cdots,M$  as a warm start at time $n$. Specifically, we can write the gradient of $f_n(\bD)$ in \eqref{ftD} at $\bD_{n-1}$ as 
\begin{equation}
 \frac{\partial f_n(\bD)}{\partial\bD}{\Big |}_{\bD=\bD_{n-1}}  =-2  \bP_n(\bx_n-\bD_{n-1}\ba_n)\ba_n^T, 
 \end{equation}
where $\ba_n$ is given in \eqref{coef}. Then the gradient of $F_n(\bD)$ at $\bD_{n-1}$ is given as
\begin{align*}  
 \frac{\partial F_n(\bD)}{\partial\bD}{\Big |}_{\bD=\bD_{n-1}} & = -2\sum_{t=1}^n \lambda^{n-t} \bP_t(\bx_t-\bD_{n-1}\ba_t)\ba_t^T.
\end{align*}
The Hessian for each row of $\bD$ at $\bd_m^{n-1}$ is therefore
\begin{align}
\bH_n(\bd_m^{n-1},\lambda) &= \frac{\partial^2 F_n(\bD)}{\partial \bd_m\partial\bd_m^T}{\Big |}_{\bd_m=\bd_m^{n-1}} \nonumber \\
&= 2\sum_{t=1}^n \lambda^{n-t} p_{mt}\ba_t\ba_t^T. \label{Hessian}
\end{align}
It follows that the update rule for each row $\bd_m$ given in \eqref{row-update} can be written as
\begin{equation} \label{pet1}
\bd_m^n=\bd_m^{n-1}-\bH_n(\bd_m^{n-1},\lambda)^{-1} \frac{\partial f_n(\bD)}{\partial\bd_m^{n-1}},
\end{equation}
which is equivalent to second-order stochastic gradient descent. Therefore, PETRELS converges to a stationary point of $F_n(\bD)$ \cite{bottou-bousquet-2008, bottou2008large}. Compared with first-order algorithms, PETRELS enjoys a faster convergence speed to the stationary point \cite{bottou-bousquet-2008,bottou2008large}.  

\subsection{Comparison with GROUSE} \label{comp-Grouse} 
%%\subsection{The GROUSE Algorithm}
%It aims to minimize the projection residual on the observed entries with respect to the underlying subspace at each time $n$, i.e.
%$$f_n(\bD) = \min_{\ba_n} \|\bP_n(\bx_n-\bD\ba_n)\|_2^2.$$
%It can be viewed as a first-order stochastic gradient descent on the orthogonal Grassmannian manifold 
%$$\mathcal{G}_r=\{\bD\in\mathbb{R}^{M\times r}:\bD^T\bD=\bI_r\}$$ 
%by using the previous estimate $\bD_{n-1}$ as a warm start, yielding the updating rule as 
%\begin{align} \label{update-D-grouse}
%\bD_n &= \bD_{n-1} - \Big[(\cos(\sigma\eta_n)-1)\frac{\hat{\bx}_n}{\|\hat{\bx}_n\|_2} + \nonumber \\
%&\quad\quad \sin(\sigma\eta_n)\frac{\br_n}{\|\br_n \|_2} \Big] \frac{\ba_n^T}{\|\ba_n\|_2},
%\end{align}
%where $\sigma=\|\hat{\bx}_t\|_2\|\br_t\|_2$, and $\eta_n$ is the step-size at time $n$. At each step GROUSE also alternates between coefficient estimation \eqref{coef} and subspace update \eqref{update-D-grouse}. If the step size satisfies 
%$$ \lim_{n\to\infty} \eta_n =0 \quad~\mbox{and}~\quad \sum_{t=1}^{\infty} \eta_t = \infty,$$
%then GROUSE is guaranteed to converge to a stationary point of $G_n(\bD)$. 
%However, performance is limited by the existence of ``barriers'' in the search path \cite{SET} which result in GROUSE being trapped at a local minima. We will provide numerical comparisons in Section~\ref{simulation} in the example of direction-of-arrival estimation. 
%

The GROUSE algorithm \cite{Balzano-2010} proposed by Balzano et. al.  addresses the same problem of online identification of low-rank subspace from highly incomplete information. The GROUSE method can be viewed as optimizing \eqref{optimization} for $\lambda=1$ at each time $n$ using a \textit{first-order} stochastic gradient descent on the \textit{orthogonal} Grassmannian defined as $\mathcal{G}_r=\{\bD\in\mathbb{R}^{M\times r}: \bD^T\bD=\bI_r\}$ instead of $\mathbb{R}^{M\times r}$. Thus, GROUSE aims to solve the following optimization problem,
\begin{equation}\label{optimization1}
\bD_n =  \argmin_{\bD\in\mathcal{G}_r} G_n(\bD)= \argmin_{\bD\in\mathcal{G}_r} \sum_{t=1}^n f_t(\bD).
\end{equation}

GROUSE updates the subspace estimate along the direction of $\nabla f_t(\bD)|_{\bD=\bD_{n-1}}$ on $\mathcal{G}_r$, given by
\begin{align} \label{update-D-grouse}
\bD_n &= \bD_{n-1} - \Big[(\cos(\sigma\eta_n)-1)\frac{\hat{\bx}_t}{\|\hat{\bx}_n\|_2} + \nonumber \\
&\quad\quad \sin(\sigma\eta_t)\frac{\br_t}{\|\br_t\|_2} \Big] \frac{\ba_n^T}{\|\ba_n\|_2},
\end{align}
where $\sigma=\|\hat{\bx}_t\|_2\|\br_t\|_2$, and $\eta_n$ is the step-size at time $n$. At each step GROUSE also alternates between coefficient estimation \eqref{coef} and subspace update \eqref{update-D-grouse}. Moreover, the resulting algorithm is a fast rank-one update on $\bD_{n-1}$ at each time $n$. Given that it is a first-order gradient descent algorithm, convergence to a stationary point but not global optimal is guaranteed under mild conditions on the step-size. Specifically, if the step size satisfies 
\begin{equation}\label{stepsize}
 \lim_{n\to\infty} \eta_n =0 \quad~\mbox{and}~\quad \sum_{t=1}^{\infty} \eta_t = \infty,
 \end{equation}
then GROUSE is guaranteed to converge to a stationary point of $G_n(\bD)$. However, due to the existence of ``barriers'' in the search path on the Grassmannian \cite{SET}, GROUSE may be trapped at a local minima as shown in Section \ref{simulation} in the example of direction-of-arrival estimation. Although both PETRELS and GROUSE have a tuning parameter, compared with the step-size in GROUSE, the discount factor in PETRELS is an easier parameter to tune. For example, without discounting (i.e. $\lambda=1$) PETRELS can still converge to the \textit{global} optimal given full observations as shown in Section~\ref{convergence}, while this is impossible to achieve with a first-order algorithm like GROUSE if the step size is not tuned properly to satisfy \eqref{stepsize}.

%The GROUSE method can be viewed as optimizing \eqref{optimization} for $\lambda=1$ at each time $n$ using a first-order stochastic gradient descent on the \textit{orthogonal} Grassmannian defined as $\mathcal{G}_r=\{\bD\in\mathbb{R}^{M\times r}: \bD^T\bD=\bI_r\}$ instead of $\mathbb{R}^{M\times r}$. Thus, in the GROUSE algorithm,
%\begin{equation}\label{optimization1}
%\bD_n =  \argmin_{\bD\in\mathcal{G}_r} G_n(\bD)= \argmin_{\bD\in\mathcal{G}_r} \sum_{t=1}^n f_t(\bD).
%\end{equation}
%GROUSE updates the subspace estimate along the direction of $\nabla f_t(\bD)|_{\bD=\bD_{n-1}}$ on $\mathcal{G}_r$, given by
%\begin{align} \label{update-D-grouse}
%\bD_n &= \bD_{n-1} - \Big[(\cos(\sigma\eta_n)-1)\frac{\hat{\bx}_t}{\|\hat{\bx}_n\|_2} + \nonumber \\
%&\quad\quad \sin(\sigma\eta_t)\frac{\br_t}{\|\br_t\|_2} \Big] \frac{\ba_n^T}{\|\ba_n\|_2},
%\end{align}
%where $\sigma=\|\hat{\bx}_t\|_2\|\br_t\|_2$, and $\eta_n$ is the step-size at time $n$. At each step GROUSE also alternates between coefficient estimation \eqref{coef} and subspace update \eqref{update-D-grouse}. If the step size satisfies 
%$$ \lim_{n\to\infty} \eta_n =0 \quad~\mbox{and}~\quad \sum_{t=1}^{\infty} \eta_t = \infty,$$
%then GROUSE is guaranteed to converge to a stationary point of $G_n(\bD)$. However, due to the existence of ``barriers'' in the search path on the Grassmannian \cite{SET}, GROUSE may be trapped at a local minima as shown in Section \ref{simulation} in the example of direction-of-arrival estimation. 

If we relax the objective function of GROUSE \eqref{optimization1} to all rank-$r$ subspaces $\mathbb{R}^{M\times r}$, given as  
\begin{equation} \label{nodiscounting}
\bD_n =  \argmin_{\bD\in\mathbb{R}^{M\times r}} \sum_{t=1}^n f_t(\bD),
\end{equation}
then the objective function becomes equivalent to PETRELS without discounting. It is possible to use a different formulation of second-order stochastic gradient descent with step-size to solve \eqref{nodiscounting}, yielding the update rule for each row of $\bD_n$ as
\begin{align} \label{second-grouse}
\bd_m^n &=\bd_m^{n-1}-\gamma_n \bH_n(\bd_m^{n-1},\lambda=1)^{-1}  \frac{\partial f_n(\bD)}{\partial\bd_m^{n-1}} ,  
\end{align}
where $\bH_n(\bd_m^{n-1},\lambda=1)$ is given in \eqref{Hessian}, and $\gamma_n$ is the step-size at time $n$. Compared with the update rule for PETRELS in \eqref{pet1}, the discount parameter has a similar role as the step-size, but weights the contribution of previous data input geometrically. However, in this paper we didn't investigate the performance of this alternative update rule in \eqref{second-grouse}.

\subsection{Complexity Issues}
We compare both storage complexity and computational complexity for PETRELS, GROUSE and the PAST algorithm. The storage complexity of PAST and GROUSE is $\mathcal{O}(Mr)$, which is the size of the low-rank subspace. On the other hand, PETRELS has a larger storage complexity of $\mathcal{O}(Mr^2)$, which is the total size of $\bR_m^n$'s for each row. In terms of computational complexity, PAST has a complexity of $\mathcal{O}(Mr)$, while PETRELS and GROUSE have a similar complexity on the order of $\mathcal{O}(|\Omega_t| r^2)$, where the main complexity comes from computation of the coefficient \eqref{coef}. This indicates another merit of dealing with partial observations, i.e. to reduce computational complexity when the dimension is high.

\section{Extensions of the PETRELS Algorithm } \label{discussions}
\subsection{Simplified PETRELS}
In the subspace update step of PETRELS in \eqref{parD}, consider replacing the objective function in \eqref{optimization} by
\begin{align} \label{subspace-update-mod}
\bD_n &=\argmin_{\bD}\hat{F}_n(\bD) \nonumber\\
&=\argmin_{\bD}\sum_{t=1}^n\lambda^{n-t}\|\hat{\bx}_t-\bD\ba_t\|_2^2,
\end{align}
where $\ba_t$ and $\hat{\bx}_t$, $t=1,\cdots, n$ are estimates from earlier steps in \eqref{coef} and \eqref{data_recon}. The only change we made is to remove the partial observation operator from the objective function, and replace it by the full vector estimate. It remains true that $\bd_m^n=\argmin_{\bd_m} \hat{F}_n(\bd_m) =\bd_m^{n-1}$ if the corresponding $m$th entry of $\bx_n$ is unobserved, i.e. $m\notin \Omega_{n}$, since
\begin{align*}
\hat{F}_n(\bd_m)& = \sum_{t=1}^{n-1}\lambda^{n-t}\|\hat{x}_{mt}-\bd_m^T\ba_t\|_2^2 + \|(\bd_m^{n-1}-\bd_m)^T\ba_t\|_2^2, \\
& = \lambda \hat{F}_{n-1}(\bd_m) +\|(\bd_m^{n-1}-\bd_m)^T\ba_t\|_2^2 \\
& \geq \lambda \hat{F}_{n-1}(\bd_m^{n-1}) = \hat{F}_{n}(\bd_m^{n-1})
\end{align*}
is minimized when $\bd_m=\bd_m^{n-1}$ for $m\notin\Omega_n$.

This modification leads to a simplified update rule for $\bR_m^n$, since now the updating formula for all rows $\bd_m$'s is the same, where $\bR_m^n=\bR_n=\lambda\bR_{n-1}+\ba_n\ba_n^T$ for all $m$. The row updating formula \eqref{row-update} is replaced by
\begin{equation} \label{joint-update}
\bD_n = \bD_{n-1}+\bP_n(\bx_n-\bD_{n-1}\ba_n)\ba_n^T\bR_n^{\dag}, 
\end{equation}
which further saves storage requirement for the PETRELS algorithm from $\mathcal{O}(Mr^2)$, to $\mathcal{O}(Mr)$ which is the size of the subspace. We compared the performance of the simplified PETRELS against PETRELS in Section~\ref{simulation}, which converges slower than PETRELS but might have an advantage if the subspace rank is underestimated.

\subsection{Incorporating Prior Information} \label{mcsimilar}
It is possible to incorporate regularization terms into PETRELS to encode prior information about the data stream. Here we outline the regularization on the subspace $\bD$ in the \textit{subspace update} step, such that at each time $n$, $\bD_n$ is updated via 
\begin{equation}\label{regD}
\bD_n=\argmin_{\bD}\sum_{t=1}^n\lambda^{n-t}\|\bP_t (\bx_t-\bD\ba_t)\|_2^2 + \mu_n \|\bD\|_F^2,
\end{equation}
where $\mu_n > 0$ is the regularization parameter. Similar as PETRELS, \eqref{regD} can be decomposed for each row of $\bD=[\bd_1,\bd_2,\cdots,\bd_M]^T$ as
\begin{align}
\bd_m^n &=\argmin_{\bd_m}\sum_{t=1}^n\lambda^{n-t}p_{mt} (x_{mt}-\ba_t^T\bd_m)^2+\mu_n \|\bd_m\|_2^2 \nonumber \\
&= \left( \sum_{t=1}^n \lambda^{n-t}p_{mt}\ba_t\ba_t^T+\mu_n \bI\right)^{-1} \left( \sum_{t=1}^n \lambda^{n-t}p_{mt}x_{mt}\ba_t\right) \nonumber \\
& =  (\bT_m^n)^{-1}\bs_m^n. \nonumber
\end{align}
The matrix $\bT_m^n$ can be updated as
\begin{align*}
\bT_m^n &= \lambda \bT_m^{n-1} + p_{mn}\ba_t\ba_t^T + (\mu_n-\lambda\mu_{n-1})\bI_r, 
\end{align*}
and $\bs_m^n$ can be updated as \eqref{rec-s}. However the fast RLS algorithm no longer applies here, so additional complexity for matrix inversion is required. It is worth noticing that \eqref{regD} closely resembles the matrix completion formula \eqref{MF} when $\bV$ is fixed and composed of columns of $\ba_t$'s.
%
%\subsection{Robust PETRELS with Sparse Outliers}
%There is an increasing need of identifying and subtracting sparse outliers in the low-rank data, examples including anomalies in network data \cite{Ahmed_Coates_Lakhina_2007} and separating backgrounds and foregrounds in video surveillance \cite{video_surv}. The data model in \eqref{model_data} is modified to incorporate sparse outliers, given as
%\begin{equation}
%\bx_t = \bU_t \ba_t + \bn_t + \bs_t,
%\end{equation}
%where $\bs_t$ is a sparse vector. We can modify the loss function to $\ell_1$ minimization penalty as
%$$ f_t(\bD) = \min_{\ba} \|\bP_t(\bx_t-\bD\ba)\|_1,\;t=1,\cdots, n. $$
%The coefficient update in \eqref{coef} by
%\begin{equation}
%\{\ba_n,\bs_n\}  =\argmin_{\ba,\bs} \|\bP_n(\bx_n-\bs_n-\bD_{n-1}\ba)\|_2^2 + \mu \|\bs_n|\|_1,
%\end{equation}

\section{Global Convergence With Full Observation} 
\label{convergence}
In the partial observation regime, the PETRELS algorithm always converges to a stationary point of $F_n(\bD)$, given it's a second-order stochastic gradient descent method in Section~\ref{second-order}, but whether it converges to the global optimal remains open. However, in the full observation regime, i.e. $\by_n=\bx_n$ for all $n$, we can show that the PETRELS algorithm converge globally as below. 

In this case, PETRELS becomes essentially equivalent to the conventional PAST algorithm \cite{Yang-1995} for SIT except that the coefficient is estimated differently. Specifically, in PAST it is estimated as $\ba_n = \bD_{n-1}^T\by_n = \bD_{n-1}^T\bx_n$, while in PETRELS it is estimated as $\ba_n =(\bD_{n-1}^T\bD_{n-1})^{-1}\bD_{n-1}^T\bx_n$. 

Now let $\lambda=1$, similar to PAST in \cite{Yang-1996}, the asymptotic dynamics of the PETRELS algorithm can be described by the ODE below, 
\begin{align}
\dot{\bR} & =  \mathbb{E}[\tilde{\ba}_n\tilde{\ba}_n^T] -\bR \nonumber \\
&= (\bD^T\bD)^{-1}\bD^T\bC_{\bx}\bD(\bD^T\bD)^{-1}- \bR, \label{dynR} \\
\dot{\bD} & = \mathbb{E}[\bx_n(\bx_n-\bD\tilde{\ba}_n)^T] \bR^{\dag} \nonumber \\
&= (\bI - \bD (\bD^T\bD)^{-1}\bD^T)\bC_{\bx}\bD (\bD^T\bD)^{-1}\bR^{-1}. \label{dynD}
\end{align}
Here $\tilde{\ba}_n =(\bD^T\bD)^{-1}\bD^T\bx_n$, $\bR=\bR(t)$ and $\bD=\bD(t)$ are continuous-time versions of $\bR_n$ and $\bD_n$. Now let $\widetilde{\bD}=\bD(\bD^T\bD)^{-1/2}$ and $\widetilde{\bR}=(\bD^T\bD)^{1/2}\bR(\bD^T\bD)^{1/2}$. From \eqref{dynD},
$$\bD^T\dot{\bD} = \bD^T (\bI - \bD (\bD^T\bD)^{-1}\bD^T)\bC_{\bx}\bD (\bD^T\bD)^{-1}\bR^{-1} = \bf{0}, $$
and
$$\frac{d}{dt}(\bD^T\bD)=\bD^T\dot{\bD} + \dot{\bD}^T\bD=\bf{0},$$
furthermore  
$$\frac{d}{dt}f(\bD^T\bD)=\bf{0}$$ for any function of $\bD^T\bD$. Hence,
\begin{align*}
\dot{\widetilde{\bD}} &=\dot{\bD}(\bD^T\bD)^{-1/2} + \bD\frac{d}{dt}(\bD^T\bD)^{-1/2} = \dot{\bD}(\bD^T\bD)^{-1/2}, 
\end{align*}
and
\begin{align*}
\dot{\widetilde{\bR}}  &=\frac{d}{dt}(\bD^T\bD)^{-1/2}\bR(\bD^T\bD)^{1/2}+(\bD^T\bD)^{1/2}\dot{\bR}(\bD^T\bD)^{1/2}\nonumber \\
&\;+(\bD^T\bD)^{1/2}\bR \frac{d}{dt}(\bD^T\bD)^{1/2} =(\bD^T\bD)^{1/2}\dot{\bR}(\bD^T\bD)^{1/2}. 
\end{align*}
Therefore \eqref{dynR} and \eqref{dynD} can be rewritten as
\begin{align*}
\dot{\widetilde{\bR}} &= \widetilde{\bD}^T \bC_{\bx} \widetilde{\bD} - \widetilde{\bR}, \\
\dot{\widetilde{\bD}} & = (\bI - \widetilde{\bD}\widetilde{\bD}^T)\bC_{\bx}\widetilde{\bD}\widetilde{\bR}^{\dag},
\end{align*} 
which is equivalent to the ODE of PAST. Hence we conclude that PETRELS will converge to the global optima in the same dynamic as the PAST algorithm.

\section{Numerical Results} \label{simulation}
Our numerical results fall into four parts. First we examine the influence of parameters specified in the PETRELS algorithm, such as discount factor, rank estimation, and its robustness to noise level. Next we look at the problem of direction-of-arrival estimation and show PETRELS demonstrates performance superior to GROUSE by identifying and tracking all the targets almost perfectly even in low SNR. Thirdly, we compare our approach with matrix completion, and show that PETRELS is at least competitive with state of the art batch algorithms. Finally, we provide numerical simulations for the extensions of the PETRELS algorithm.

\subsection{Choice of Parameters}
At each time $t$, a vector $\bx_t$ is generated as
\begin{equation}
\bx_t = \bD_{true} \ba_t +\bn_t, t=1,2,\cdots
\end{equation}
where $\bD_{true}$ is an $r$-dimensional subspace generated with i.i.d. $\mathcal{N}(0,1)$ entries, $\ba_t$ is an $r\times 1$ vector with i.i.d. $\mathcal{N}(0,1)$ entries, and $\bn_t$ is an $m\times 1$ Gaussian noise vector with i.i.d. $\mathcal{N}(0,\epsilon^2)$ entries. We further fix the signal dimension $m=500$ and the subspace rank $r_{true}=10$. We assume that a fixed number of entries in $\bx_t$, denoted by $K$, are revealed each time. This restriction is not necessary for the algorithm to work as shown in matrix completion simulations, but we make it here in order to get a meaningful estimate of $\ba_t$. Denoting the estimated subspace by $\hat{\bD}$, we use the normalized subspace reconstruction error to examine the algorithm performance. This is calculated as $\|\cP_{\hat{\bD}_{\perp}}\bD_{true}\|_F^2/\|\bD_{true}\|_F^2$, where $\cP_{\hat{\bD}_{\perp}}$ is the projection operator to the orthogonal subspace $\hat{\bD}_{\perp}$ .

\begin{figure}[htp]
\centering
\includegraphics[width=0.5\textwidth]{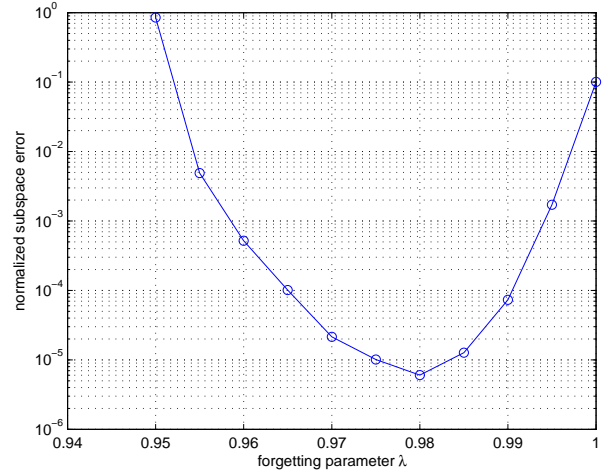} 
\caption{The normalized subspace reconstruction error as a function of the discount factor $\lambda$ after running the algorithm to time $n=2000$ when $50$ out of $500$ entries of the signal are observed each time without noise.}\label{lambda}
\end{figure}

The choice of discount factor $\lambda$ plays an important role in how fast the algorithm converges. With $K=50$, a mere $10\%$ percent of the full dimension, the rank is given accurately as $r=10$ in a noise-free setting where $\epsilon=0$. We run the algorithm to time $n=2000$ for the same data, and find that the normalized subspace reconstruction error is minimized when $\lambda$ is around $0.98$ as shown in Fig.~\ref{lambda}. Hence, we will keep $\lambda=0.98$ hereafter.

\begin{figure}[htp]
\centering
\includegraphics[width=0.5\textwidth]{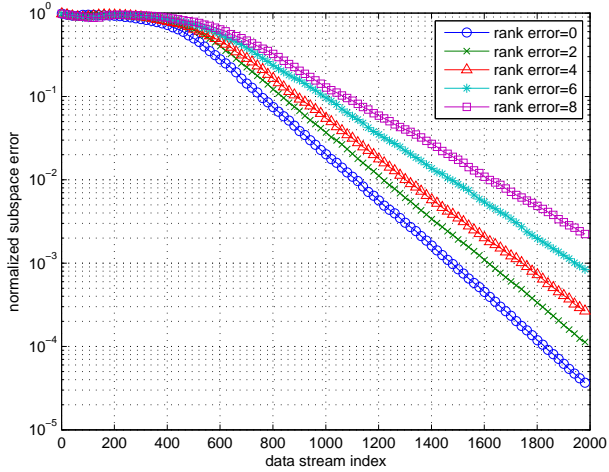} 
\caption{Normalized subspace reconstruction error as a function of data stream index when the rank is over-estimated when $50$ out of $500$ entries of the signal are observed each time without noise.}\label{par_suberr_varrank}
\end{figure}

In reality it is almost impossible to accurately estimate the intrinsic rank in advance. Fortunately the convergence rate of our algorithm degrades gracefully as the rank estimation error increases. In Fig.~\ref{par_suberr_varrank}, the evolution of normalized subspace error is plotted against data stream index, for rank estimation $r=10,12,14,16,18$. We only examine over-estimation of the rank here since this is usually the case in applications. In the next section we show examples for the case of rank underestimation.
 
\begin{figure}[htp]
\centering
\includegraphics[width=0.5\textwidth]{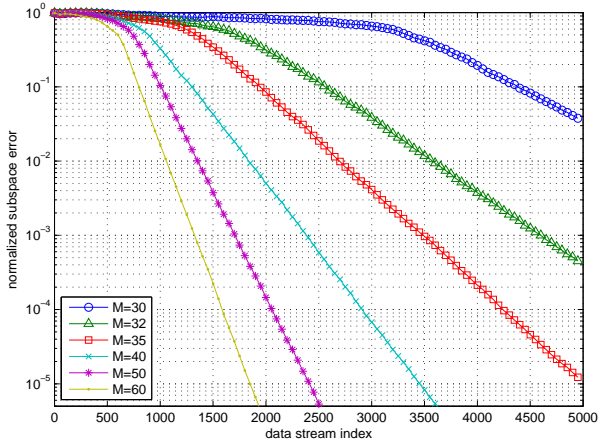} 
\caption{Normalized subspace reconstruction error as a function of data stream index when the number of entries observed per time $M$ out of $500$ entries are varied with accurate rank estimation and no noise.} \label{varM}
\end{figure}

Taking more measurements per time leads to faster convergence since it is approaching the full information regime, as shown in Fig.~\ref{varM}. Theoretically it requires $M\sim\mathcal{O}(r\log r)\approx 23$ measurements to test if an incomplete vector is within a subspace of rank $r$ \cite{Balzano-subspace}. The simulation shows our algorithm can work even when $M$ is close to this lower bound.

\begin{figure}[htp]
\centering
\includegraphics[width=0.5\textwidth]{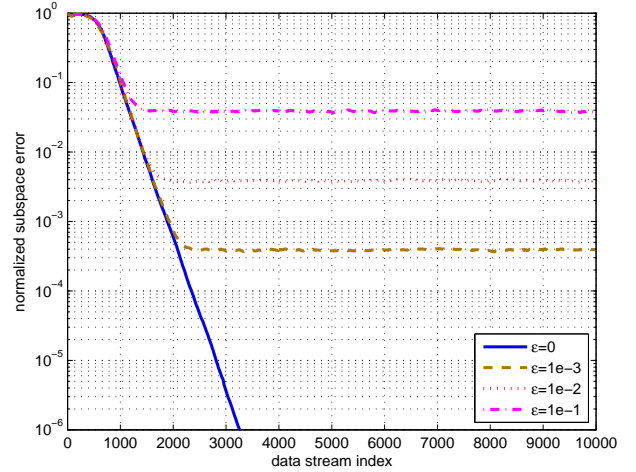} 
\caption{Normalized subspace error against data stream index with different noise level $\epsilon$  when $50$ out of $500$ entries of the signal are observed each time with accurate rank estimation.} \label{par_suberr_noise} 
\end{figure}

Finally the robustness of the algorithm is tested against the noise variance $\epsilon^2$ in Fig.~\ref{par_suberr_noise}, where the normalized subspace error is plotted against data stream index for different noise levels $\epsilon$. The estimated subspace deviates from the ground truth as we increase the noise level, hence the normalized subspace error degrades gracefully and converges to an error floor determined by the noise variance. 

\begin{figure*}[htp]
\centering
 \begin{tabular}{cc}
\includegraphics[width=0.5\textwidth]{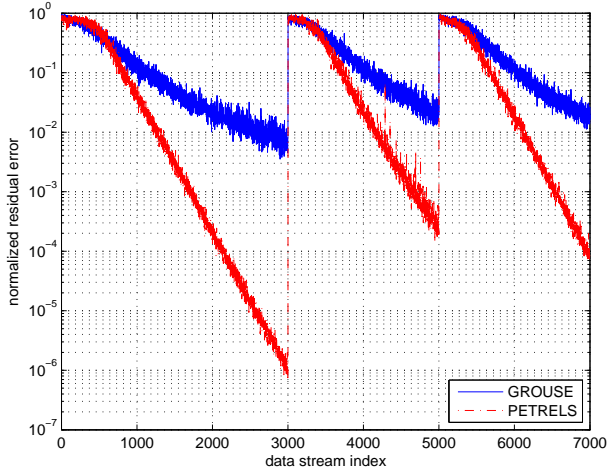} & 
\includegraphics[width=0.5\textwidth]{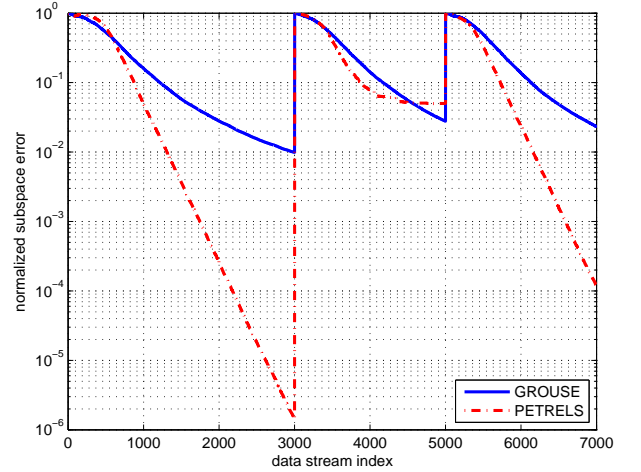} \\
(a) Normalized residual error & (b) Normalized subspace error
\end{tabular}
\caption{The normalized subspace error when the underlying subspace is changing with fixed rank $r=10$. The rank is over-estimated by $4$ and the noise level is $\epsilon=10^{-3}$, when $50$ out of $500$ entries of the signal are observed each time for both GROUSE and PETRELS.} \label{tracking} 
\end{figure*}

We now consider a scenario where a subspace of rank $r=10$ changes abruptly at time index $n=3000$ and $n=5000$, and examine the performance of GROUSE \cite{Balzano-2010} and PETRELS in Fig.~\ref{tracking} when the rank is over-estimated by $4$ and the noise level is $\epsilon=10^{-3}$. The normalized residual error for data stream, calculated as $\|\bP_n(\bx_n-\hat{\bx}_n)\|_2/\|\bP_n\bx_n\|_2$, is shown in Fig.~\ref{tracking} (a), and the normalized subspace error is shown in Fig.~\ref{tracking} (b) respectively. Both PETRELS and GROUSE can successfully track the changed subspace, but PETRELS can track the change faster.

\subsection{Direction-Of-Arrival Analysis}
Given GROUSE \cite{Balzano-2010} as a baseline,  we evaluate the resilience of our algorithm to different data models and applications. We use the following example of Direction-Of-Arrival analysis in array processing to compare the performance of these two methods. Assume there are $n=256$ sensors from a linear array, and the measurements from all sensors at time $t$ are given as
\begin{equation}
\bx_t = \bV\bm{\Sigma}\ba_t + \bn_t, \quad t=1,2,\cdots.
\end{equation}
Here $\bV\in\mathbb{C}^{n\times p}$ is a Vandermonde matrix given by
\begin{equation}
\bV= [\bm{\alpha}_1(\omega_1), \cdots, \balpha_p(\omega_p)],
\end{equation}
where $\balpha_i(\omega_i) = [1, e^{j2\pi\omega_i},\cdots,e^{j2\pi\omega_i(n-1)]}]^T$, $0\leq\omega_i<1$; $\bm{\Sigma}=\mbox{diag}\{\bd\}=\mbox{diag}\{d_1,\cdots,d_p\}$ is a diagonal matrix which characterizes the amplitudes of each mode. The coefficients $\ba_t$ are generated with $\mathcal{N}(0,1)$ entries, and the noise is generated with $\mathcal{N}(0,\epsilon^2)$ entries, where $\epsilon=0.1$. 

Each time we collect measurements from $K=30$ random sensors. We are interested in identifying all $\{\omega_i\}_{i=1}^p$ and $\{d_i\}_{i=1}^p$. This can be done by applying  the well-known ESPRIT algorithm \cite{Kailath-ESPRIT} to the estimated subspace $\hat{\bD}$ of rank $r$, where $r$ is specified a-priori corresponding to the number of modes to be estimated. Specifically, if $\bD_1=\hat{\bD}(1:n-1)$ and $\bD_2=\hat{\bD}(2:n)$ are the first and the last $n-1$ rows of $\hat{\bD}$, then from the eigenvalues of the matrix $\mathbf{T}=\bD_1^{\dag}\bD_2$, denoted by $\lambda_i$, $i=1,\cdots, r$, the set of $\{\omega_i\}_{i=1}^p$ can be recovered as
\begin{equation}
\omega_i = \frac{1}{2\pi}\arg{\lambda_i}, \; i=1,\cdots, r.
\end{equation}
The ESPRIT algorithm also plays a role in recovery of multi-path delays from low-rate samples of the channel output \cite{Eldar-PS}.

We show that in a dynamic setting when the underlying subspace is varying, PETRELS does a better job of discarding out-of-date modes and picking up new ones in comparison with GROUSE. We divide the running time into $4$ parts, and the frequencies and amplitudes are specified as follows:
\begin{enumerate}
\item Start with the same frequencies 
$$\omega = [0.1769,  \;  0.1992,   \; 0.2116, \;    0.6776,\;    0.7599];$$
and amplitudes
$$ d = [ 0.3,  \;  0.8,  \;   0.5,  \;  1,  \;  0.1]. $$
\item Change two modes (only frequencies) at stream index $1000$:
$$\omega = [0.1769,  \;  0.1992,   \; \mathbf{0.4116}, \;    0.6776,\;   \mathbf{ 0.8599}];$$
and amplitudes
$$ d = [ 0.3,  \;  0.8,  \;   0.5,  \;  1,  \;  0.1]. $$
\item Add one new mode at stream index $2000$:
$$\omega = [0.1769,  0.1992,   0.4116,  0.6776, 0.8599, \mathbf{0.9513}];$$
and amplitudes
$$ d = [ 0.3,  \;  0.8,  \;   0.5,  \;  1,  \;  0.1, \mathbf{0.6}]. $$
\item Delete the weakest mode at stream index $3000$:
$$\omega = [0.1769,  \;  0.1992,   \; 0.4116, \;    0.6776,\;    0.9513];$$
and amplitudes
$$ d = [ 0.3,  \;  0.8,  \;   0.5,  \;  1,   \; 0.6]. $$
\end{enumerate}

\begin{figure}[htp]
\centering
\includegraphics[width=0.5\textwidth, viewport=0 0 450 315, clip=true]{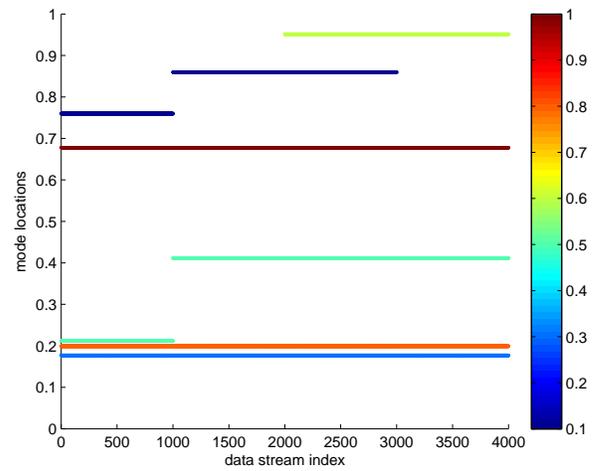} 
\caption{Ground truth of the actual mode locations and amplitudes in a dynamic scenario.} \label{groundtruth}
\end{figure}

Fig.~\ref{groundtruth} shows the ground truth of mode locations and amplitudes for the scenario above. Note that there are three closely located modes and one weak mode in the beginning, which makes the task challenging. We compare the performance of PETRELS and GROUSE. The rank specified in both algorithms is $r=10$, which is the number of estimated modes at each time index; in our case it is twice the number of true modes.\footnote{In practice the number of modes can be estimated via the Maximum Description Length (MDL) algorithm \cite{rissanen1983universal}.} 

\begin{figure*}[htp]
\centering
\begin{tabular}{cc}
\includegraphics[width=0.5\textwidth, viewport=0 0 450 315, clip=true]{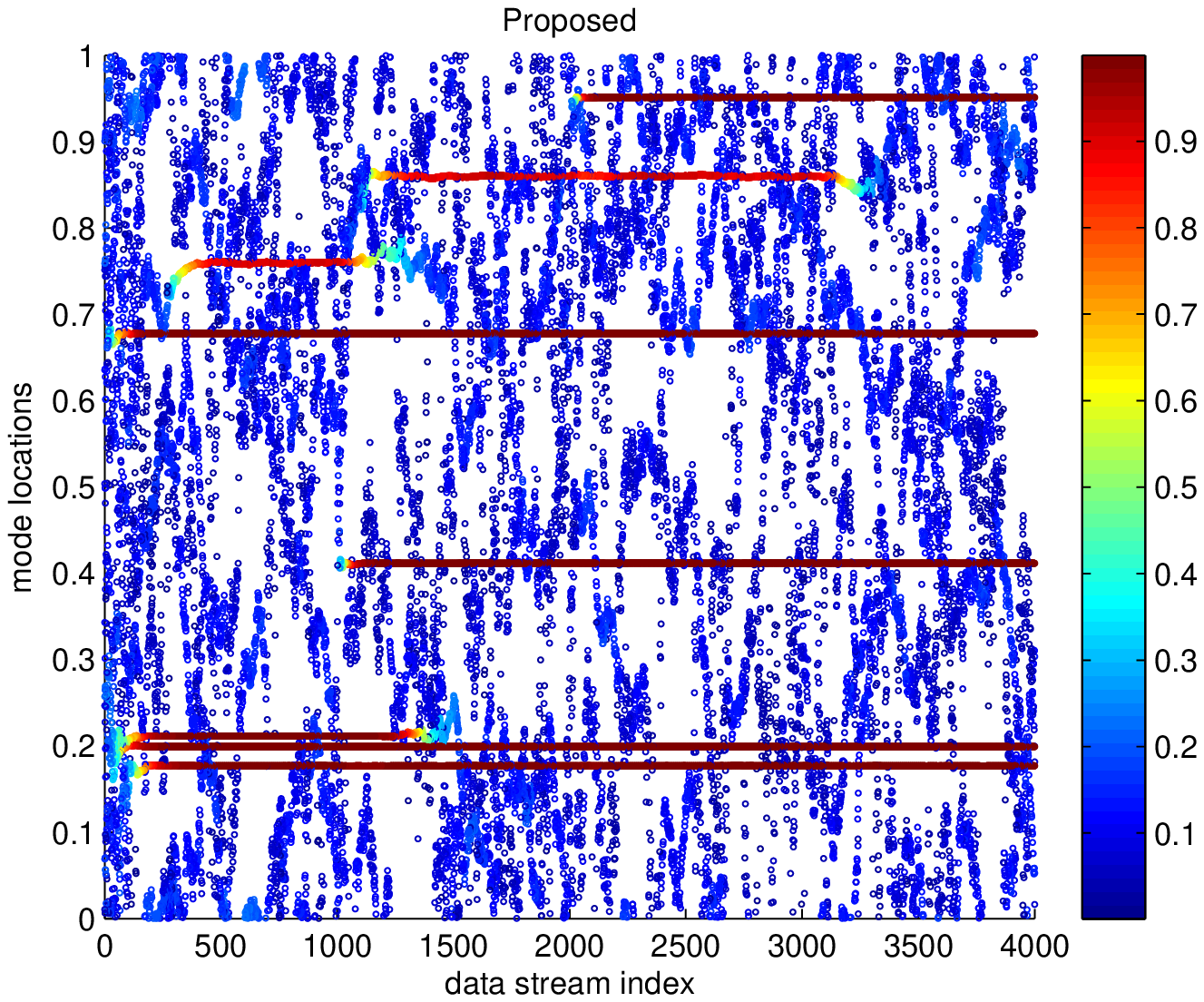} & 
\includegraphics[width=0.5\textwidth, viewport=0 0 450 315, clip=true]{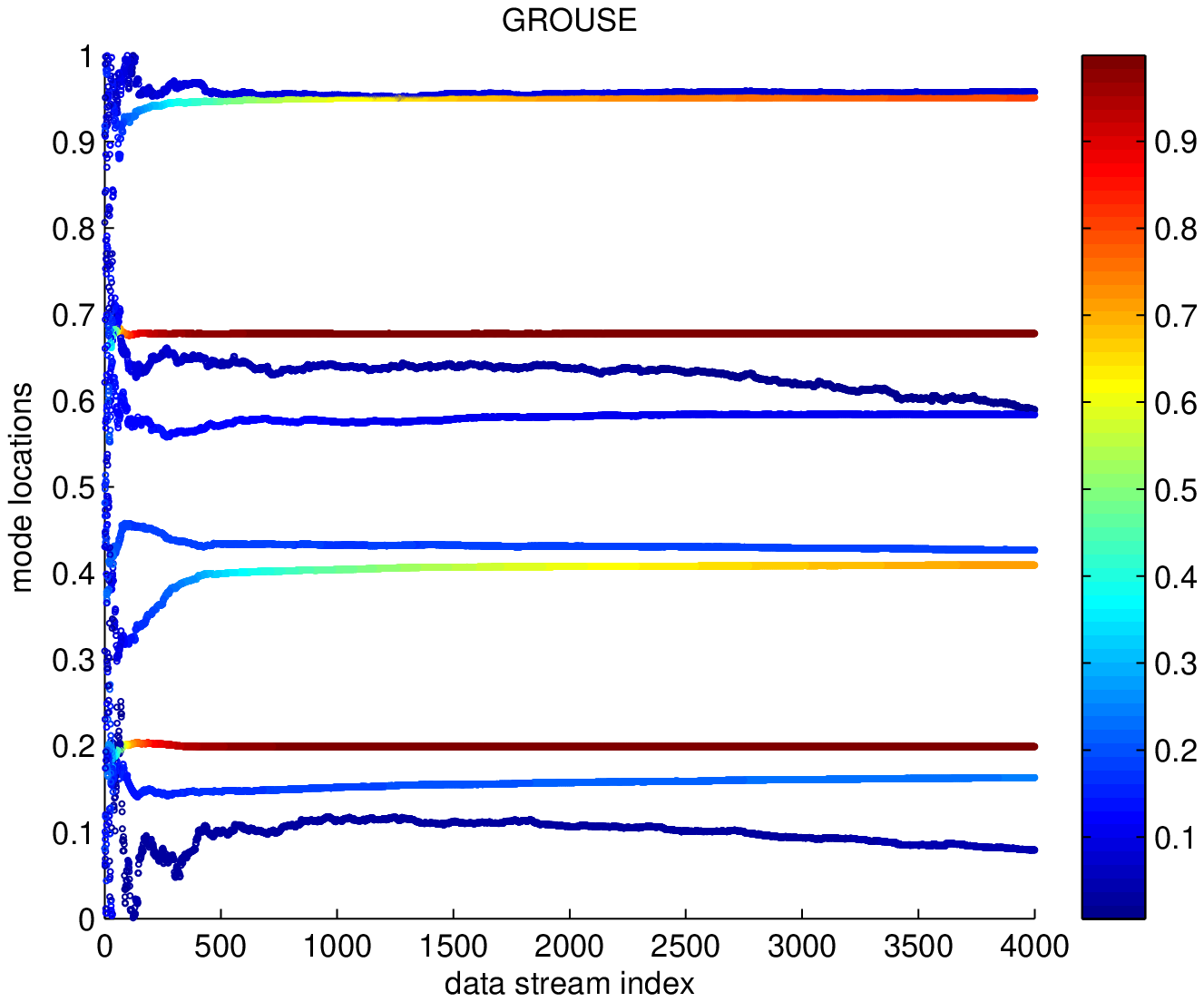} \\
(a) PETRELS & (b) GROUSE \\
\includegraphics[width=0.5\textwidth, viewport=0 0 450 315, clip=true]{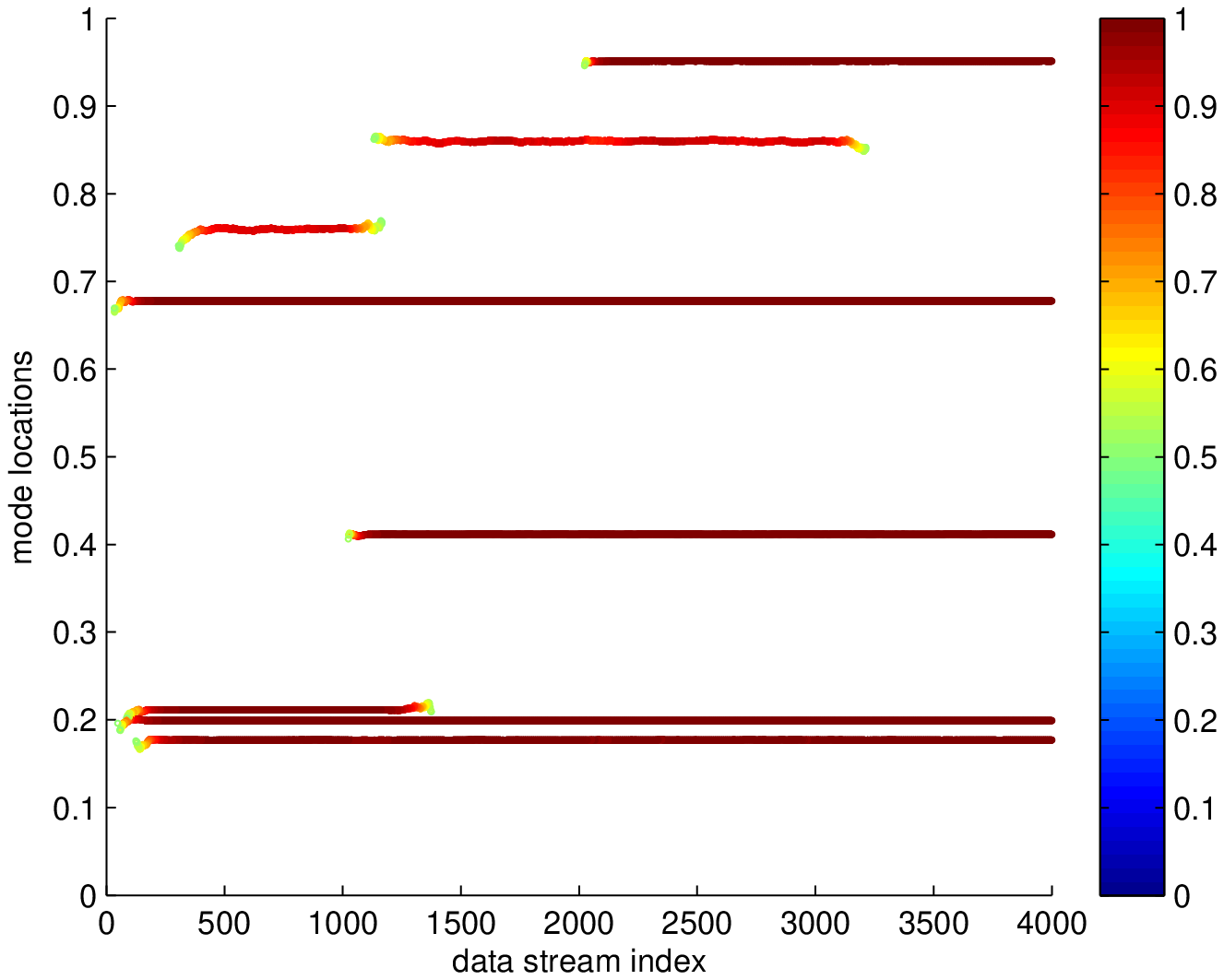} & 
\includegraphics[width=0.5\textwidth, viewport=0 0 450 315, clip=true]{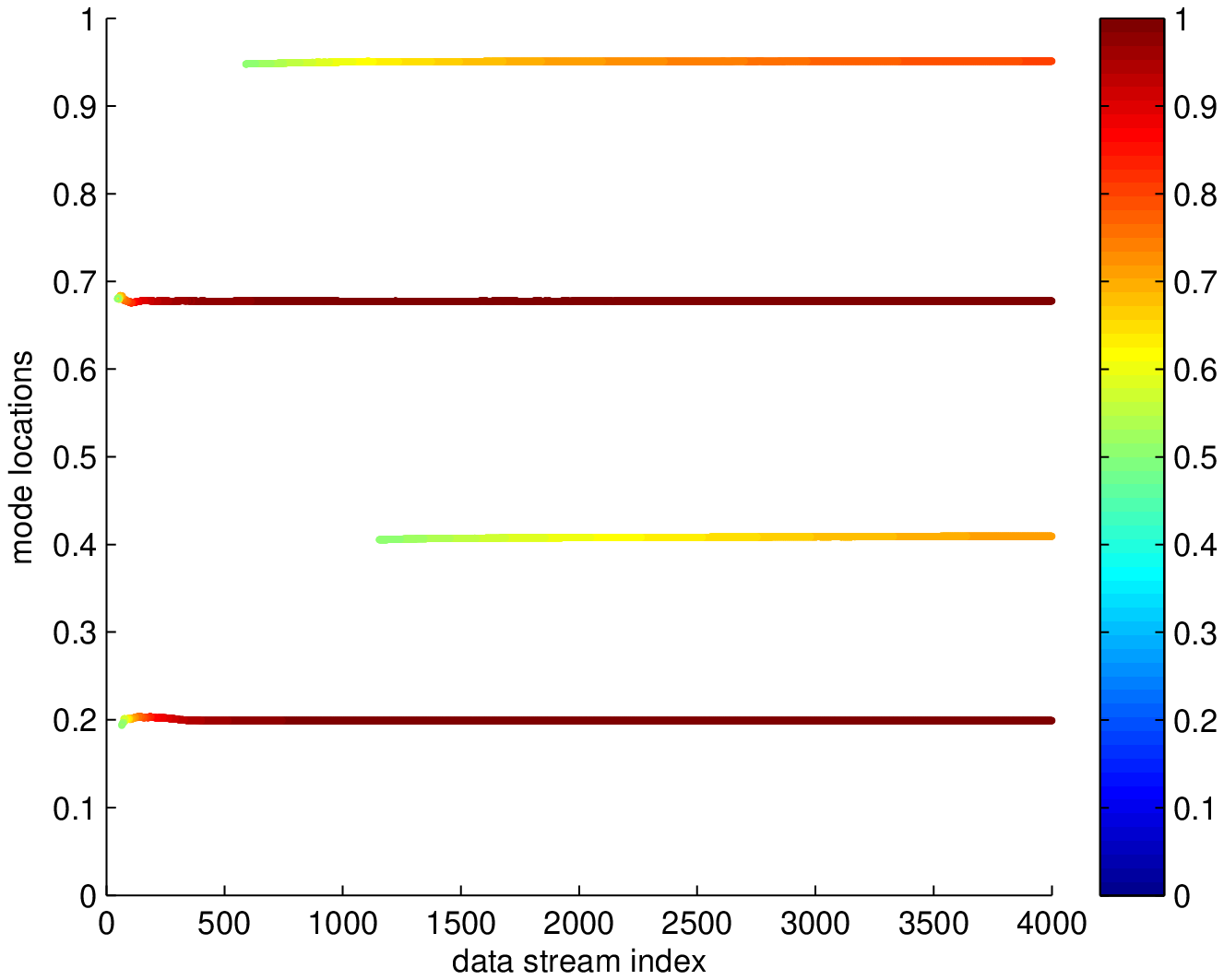} \\
(c)  PETRELS (thresholded) & (d)  GROUSE (thresholded)
\end{tabular}
\caption{Tracking of mode changes in direction-of-arrival estimation using PETRELS and GROUSE algorithms: the estimated directions at each time for $10$ modes are shown against the data stream in (a) and (b) for PETRELS and GROUSE respectively. The estimations in (a) and (b) are further thresholded with respect to level $0.5$, and the thresholded results are shown in (c) and (d) respectively. All changes are identified and tracked successfully by PETRELS, but not by GROUSE. } \label{trackmode1}
\end{figure*}

Each time both algorithms estimated $10$ modes, with their amplitude shown shown against the data stream index in Fig.~\ref{trackmode1} (a) and (b). The color shows the amplitude corresponding to the color bar. The direction-of-arrival estimations in Fig.~\ref{trackmode1} (a) and (b) are further thresholded with respect to level $0.5$, and the thresholded results are shown in Fig.~\ref{trackmode1} (c) and (d) for PETRELS and GROUSE respectively. PETRELS identifies all modes correctly. In particular PETRELS distinguishes the three closely-spaced modes perfectly in the beginning, and identifies the weak modes that come in the scene at a later time. With GROUSE the closely spaced nodes are erroneously estimated as one mode, the weak mode is missing, and spurious modes have been introduced. PETRELS also fully tracked the later changes in accordance with the entrance and exit of each mode, while GROUSE is not able to react to changes in the data model.

Since the number of estimated modes at each time is greater than the number of true modes, the additional rank in the estimated subspace contributes ``auxiliary modes'' that do not belong to the data model. In PETRELS these modes exhibit as scatter points with small amplitudes as in Fig.~\ref{trackmode1} (a), so they will not be identified as actual targets in the scene. While in GROUSE these auxiliary modes are tracked and appear as spurious modes. All changes are identified and tracked successfully by PETRELS, but not by GROUSE.

\subsection{Matrix Completion}
We next compare performance of PETRELS for matrix completion against batch algorithms LMaFit \cite{LMafit}, FPCA \cite{FPCA}, Singular Value Thresholding (SVT) \cite{SVT}, OptSpace \cite{OptSpace} and GROUSE \cite{Balzano-2010}. The low-rank matrix is generated from a matrix factorization model with $\bX=\bU\bV^T\in\mathbb{R}^{1000\times 2000}$, where $\bU\in\mathbb{R}^{1000\times 10}$ and $\bV\in\mathbb{R}^{2000\times 10}$, all entries in $\bU$ and $\bV$ are generated from standard normal distribution $\cN(0,1)$ (Gaussian data) or uniform distribution $\mathcal{U}[0,1]$ (uniform data). The sampling rate is taken to be $0.05$, so only $5\%$ of all entries are revealed.

\begin{figure*}[htp]
\centering
\begin{tabular}{cc}
\includegraphics[width=0.45\textwidth]{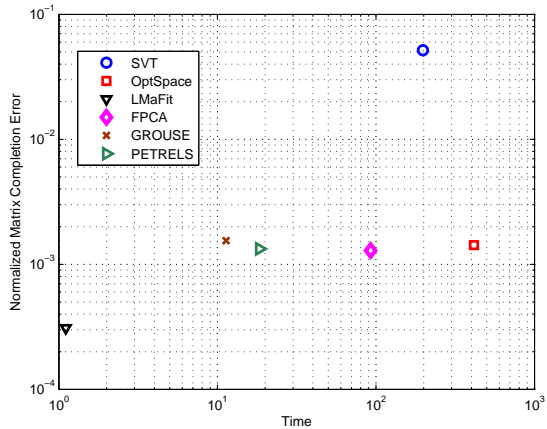} & 
\includegraphics[width=0.45\textwidth]{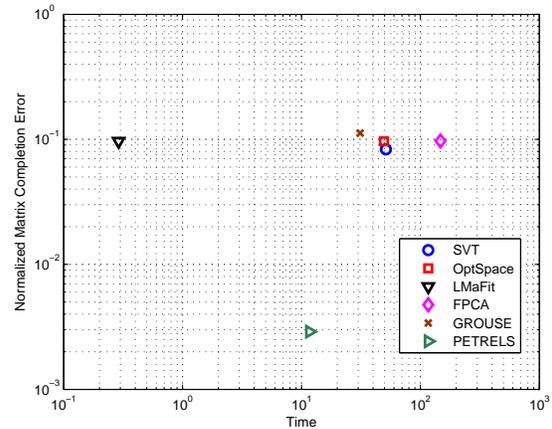} \\
(a) matrix factor from $\cN(0,1)$ & (b) matrix factor from $\mathcal{U}[0,1]$
\end{tabular}
\caption{Comparison of matrix completion algorithms in terms of speed and accuracy: PETRELS is a competitive alternative for matrix completion tasks.} \label{mc}
\end{figure*}

The running time is plotted against the normalized matrix reconstruction error, calculated as $\|\hat{\bX}-\bX\|_F/\|\bX\|_F$, where $\hat{\bX}$ is the reconstructed low-rank matrix for Gaussian data and uniform data respectively in Fig.~\ref{mc} (a) and (b). PETRELS matches the performance of batch algorithms on Gaussian data and improves upon the accuracy of most algorithms on uniform data, where the Grassmaniann-based optimization approach may encounter ``barriers'' for its convergence. Note that different algorithms have different input parameter requirements. For example, OptSpace needs to specify the tolerance to terminate the iterations, which directly decides the trade-off between accuracy and running time; PETRELS and GROUSE require an initial estimate of the rank. Our simulation here only shows one particular realization and we simply conclude that PETRELS is competitive. 

\subsection{Simplified PETRELS} 
Under the same simulation setup as for Fig.~\ref{par_suberr_varrank} except that the subspace of rank $10$ is generated by $\hat{\bD}_{true}=\bD_{true}\mathbf{\Sigma}$, where $\mathbf{\Sigma}$ is a diagonal matrix with $5$ entries from $\cN(0,1)$ and $5$ entries  from $0.01\cdot\cN(0,1)$, we examine the performance of the simplified PETRELS algorithm (with optimized $\lambda=0.9$) in Section~\ref{discussions} A and the original PETRELS (with $\lambda=0.98$) algorithm when the rank of the subspace is over-estimated as $12$ or under-estimated as $8$. When the rank of $\bD$ is over-estimated, the change in \eqref{subspace-update-mod} will introduce more errors and converges slower compared with the original PETRELS algorithm; however, when the rank of $\bD$ is under-estimated, the simplified PETRELS performs better than PETRELS. This is an interesting feature of the proposed simplification, and quantitative justification of this phenomenon is beyond the scope of this paper. Intuitively, when the rank is under-estimated, the simplified PETRELS also uses the interpolated entries to update the subspace estimate, which seems to help the performance. 

 \begin{figure}[htp]
\centering
\includegraphics[width=0.5\textwidth]{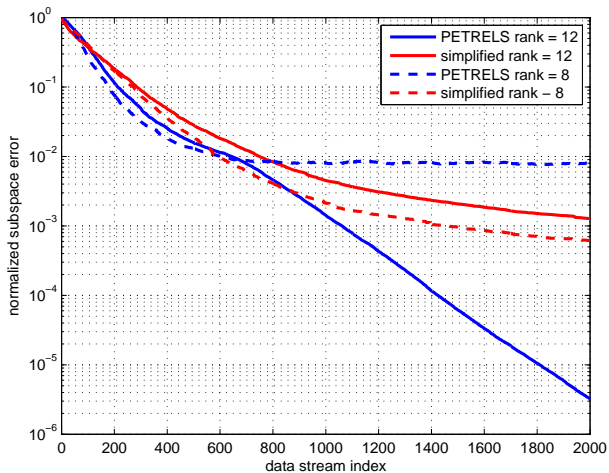} 
\caption{Normalized subspace reconstruction error against data stream index when the rank is over-estimated as $12$ or under-estimated as $8$ for the original PETRELS and modified algorithm.}\label{subspace-update-mod}
\end{figure}

\section{Conclusions} \label{conclusion}
We considered the problem of reconstructing a data stream from a small subset of its entries, where the data stream is assumed to lie in a low-dimensional linear subspace, possibly corrupted by noise. This has significant implications for lessening the storage burden and reducing complexity, as well as tracking the changes for applications such as video denoising, network monitoring and anomaly detection when the problem size is large. The well-known low-rank matrix completion problem can be viewed as a batch version of our problem. The PETRELS algorithm first identifies the underlying low-dimensional subspace via a discounted recursive procedure for each row of the subspace matrix in parallel, then reconstructs the missing entries via least-squares estimation if required. The discount factor allows the algorithm to capture long-term behavior as well as track the changes of the data stream. We show that PETRELS converges to a stationary point given it is a second-order stochastic gradient descent algorithm. In the full observation scenario, we further prove that PETRELS actually convergence globally by revealing its connection with the PAST algorithm. We demonstrate superior performance of PETRELS in direction-of-arrival estimation and showed that it is competitive with state of the art batch matrix completion algorithms.

\bibliographystyle{IEEEtran}	% (uses file "plain.bst")
\bibliography{subspace}		% expects file "myrefs.bib"

\end{document}